\documentclass[aps,prb,twocolumn,superscriptaddress,showpacs]{revtex4}

\usepackage{amsmath,amssymb}
\usepackage{graphicx,dcolumn,bm}
\usepackage[hypertex,breaklinks=true,colorlinks=false]{hyperref}

\newcommand{\Tr}{\mathop{\mathrm{Tr}}}

\newcommand{\topo}{\mathrm{topo}}
\newcommand{\Rin}{R_{\mathrm{in}}}
\newcommand{\Rout}{R_{\mathrm{out}}}
\newcommand{\cbar}{{\bar{c}}}
\newcommand{\Omegabar}{{\bar{\Omega}}}

\newcommand{\Dcal}{{\cal D}}
\newcommand{\Ecal}{{\cal E}}
\newcommand{\bra}[1]{\langle #1|}
\newcommand{\ket}[1]{|#1 \rangle}
\newcommand{\bracket}[2]{\langle #1|#2 \rangle}

\newcommand{\RK}{\mathrm{RK}}
\newcommand{\KP}{\mathrm{KP}}
\newcommand{\LW}{\mathrm{LW}}
\newcommand{\beqn}{\begin{equation}}
\newcommand{\eeqn}{\end{equation}}

\newcommand{\rhombus}{
  \put(0,0){\circle*{4}}
  \put(20,0){\circle*{4}}
  \put(10,17.3){\circle*{4}}
  \put(30,17.3){\circle*{4}}
}
\newcommand{\QDMtz}{\unitlength0.05em 
 \begin{minipage}{35\unitlength}
 \begin{center}
 \begin{picture}(30,17)
  \rhombus
  \put(0,0){\line(1,0){20}}
  \put(10,17.3){\line(1,0){20}}
 \end{picture}
 \end{center}
 \end{minipage}
}
\newcommand{\QDMts}{\unitlength0.05em 
 \begin{minipage}{35\unitlength}
 \begin{center}
 \begin{picture}(30,17)
  \rhombus
  \qbezier(0,0)(5,8.65)(10,17.3)
  \qbezier(20,0)(25,8.65)(30,17.3)
 \end{picture}
 \end{center}
 \end{minipage}
}

\begin{document}


\title{Topological Entanglement Entropy in the Quantum Dimer Model\\
on the Triangular Lattice}



\author{Shunsuke Furukawa}
\altaffiliation{Present address: Condensed Matter Theory Laboratory, RIKEN, Wako, Saitama 351-0198, Japan}
\affiliation{Department of Physics, Tokyo Institute of Technology, Meguro-ku, Tokyo 152-8551, Japan}
\affiliation{Institute for Solid State Physics, University of Tokyo, Kashiwa 277-8581, Japan}

\author{Gr\'egoire Misguich}
\affiliation{Service de Physique Th\'eorique, CEA Saclay, 91191 Gif-sur-Yvette Cedex, France}


\date{\today}

\begin{abstract}

A characterization of topological order in terms of
bi-partite
entanglement was
proposed recently [A.  Kitaev and  J.  Preskill, Phys. Rev. Lett. {\bf
96}, 110404 (2006); M. Levin  and X.-G. Wen, {\it  ibid}, 110405].  It
was argued that in a topological phase there is a universal additive
constant  in    the entanglement   entropy,  called  the {\it  topological
entanglement entropy}, which reflects the underlying gauge  theory for the 
topological order.   In the present  paper, we evaluate numerically the
topological entanglement entropy  in  the ground-states  of a quantum
dimer  model on the   triangular lattice,  which   is known to have  a
dimer liquid phase with $\mathbb{Z}_2$ topological order.    We   examine
the two original constructions to measure the  topological entropy 
by combining entropies on plural areas, 
and we observe that in the large-area limit they both  approach the value expected for $\mathbb{Z}_2$
topological   order.  We  also consider the
entanglement entropy  on a topologically non-trivial  ``zigzag'' area
and propose  to use it  as another way to measure the topological entropy.
\end{abstract}



\maketitle

\section{Introduction}
Exotic   phenomena in quantum   many-body systems  are  accompanied by
non-trivial patterns of  entanglement in  ground-state wave-functions.
One useful measure of entanglement  for a many-body state $\ket{\Psi}$
is the entanglement entropy $S_\Omega$  between a part $\Omega$ of the
system and the  rest of the system, $\Omegabar$.  It is defined  as the
von  Neumann   entropy of  the  reduced density   matrix $\rho_\Omega$
obtained by tracing out the degrees of freedom of $\Omegabar$:
\begin{equation}
 S_\Omega=-\Tr \rho_\Omega \ln \rho_\Omega, 
 \qquad \rho_\Omega=\Tr_\Omegabar \ket{\Psi}\bra{\Psi}.
\end{equation}
It has  been   clarified in the  past  few  years  that some important
properties of   a      quantum ground-state   are       encoded in the
size-dependence  of   $S_\Omega$.    For  a  system  with  short-range
correlations  only, $\Omega$ and $\Omegabar$   correlate  only in  the
vicinity of the  boundary  separating them  and thus  the entanglement
entropy  scales with   the   size of  the    boundary  ({\em  boundary
law}).\cite{Srednicki93}   However,    at  a    critical   point  with
algebraically   decaying correlations,  the  scaling of   entanglement
entropy exhibits a universal logarithmic correction characterizing the
criticality.    Specifically, in  a  one-dimensional  quantum critical
system described by  a conformal field  theory (CFT), the entanglement
entropy shows a logarithmic scaling  law with a coefficient determined
by the central charge of the CFT.\cite{Vidal03}
In some two-dimensional quantum critical states, the entanglement entropy
also contains a universal contribution, related to the geometry
of the subsystem.\cite{Fradkin06}

Another type of  non-trivial entanglement can exist  in  a system with
{\it topological order}.\cite{Wen90,Wen04}
Such   a system exhibits  degenerate  ground-states  separated from excited
states by an energy gap,
and
this   degeneracy, which depends  on the  topology   of the entire system, 
cannot be ascribed to  any type of conventional spontaneous symmetry breaking.
Indeed,  it has been  demonstrated in some models  
that these  degenerate
ground       states   cannot   be   distinguished     by     any local
observable.\cite{Ioffe02_PRB,Furukawa06,Furukawa07}     Preskill\cite{Preskill00}
suggested that this degeneracy can be regarded as a global encoding of
information reminiscent of quantum   error-correcting codes and   is a
consequence of some long-distance entanglement.  A characterization of
this global entanglement was realized  recently by Kitaev and Preskill
(KP)\cite{Kitaev06} and by  Levin and Wen (LW)\cite{Levin06}.   It was
argued that, if $\Omega$ is a  disk (in a two-dimensional system) with
a smooth boundary of length $L$, the entanglement entropy scales as
\begin{equation} \label{eq:ent_scaling}
 S_\Omega = \alpha L - \gamma + \cdots, 
\end{equation}
where the ellipsis represents terms which are  negligible in the limit
$L\to\infty$.   If  the area   $\Omega$ is  not   a disk and  has  $m$
disconnected   boundaries,    the topological     term   $-\gamma$  in
Eqn.~\eqref{eq:ent_scaling}  is    multiplied  by  $m$.   While  the
coefficient $\alpha$ depends on the microscopic details of the system,
$\gamma$ is a universal constant characterizing topological order and
was dubbed the 
{\em topological entanglement entropy}.  Indeed, $\gamma$ measures the
so-called {\em total quantum  dimension} $\Dcal$ of topological  order
by $\gamma=\ln\Dcal$. In the case of  topological order described by a
{\em discrete} Abelian gauge  theory (e.g., $\mathbb{Z}_n$), $\Dcal$ is  equal
to  the number  of elements in  the  gauge group.  In   general, it is
difficult to separate the topological term $-\gamma$ from the boundary
term in  Eqn.~\eqref{eq:ent_scaling} because,
on a lattice,  the discrete nature of the  boundary makes it difficult
to define unambiguously the length $L$.
To solve this, KP  and LW found  some ways to  define $\gamma$  by forming a
linear combination
of the entanglement entropies on plural areas sharing some boundaries,
and cancelling the  boundary terms out to  leave the topological term.
KP  and LW illustrated this idea using
effective field theories and exactly solvable models.

In  this  paper, we analyze the  entanglement  entropy  in the quantum
dimer model  (QDM)  on   the triangular  lattice\cite{Moessner01}  and
examine    the   effectiveness     of  the   proposal   in   numerical
calculations of finite-size systems. 
This model is known to exhibit a dimer liquid phase with
$\mathbb{Z}_2$ topological order in a finite interval in the parameter
space.\cite{Moessner01}  We mainly consider the Rokhsar-Kivelson (RK)
point,\cite{Rokhsar88} where the ground-states are exactly known
and  where  the  calculation of  reduced  density matrices  (and  thus
entanglement entropy) amounts to counting the number of dimer coverings
of the lattice satisfying some particular constraints.
We calculate  the  topological entanglement entropy numerically, 
and   compare the result   with $\gamma=\ln 2$
expected for $\mathbb{Z}_2$ topological order.

We  comment on related   systems here. Kitaev's model\cite{Kitaev03}  is
known to be the simplest solvable model with $\mathbb{Z}_2$ topological
order, and  the entanglement entropy  of this model  has been analyzed
rigorously  in   Refs.~\onlinecite{Hamma05,Levin06}
and the value $\gamma=\ln 2$ for the topological entropy was confirmed.
The solvable QDM (kagome lattice) of Ref.~\onlinecite{Misguich02}
can be mapped onto 
Kitaev's  model on the  honeycomb lattice,  and thus its  entanglement
entropy  can be analyzed in the  same  way.  These models give elegant
results, but are too ideal for discussing generic features
of topological order because they  have a strictly zero spin-spin  (or
dimer-dimer in the QDM) correlation  length and are completely free of
finite-size effects. In this  sense,  our analysis  on the QDM  on the
triangular lattice
is a step toward more realistic systems -- 
though we mainly   consider the exact  RK  ground-states, they have  a
finite dimer-dimer correlation length and finite-size effects arise.
In the same spirit but for another 
kind of  topological order, the entanglement  entropy of Laughlin wave
functions was analyzed numerically in Ref.~\onlinecite{Haque06}.

The paper is organized as  follows.  In Sec.~\ref{sec:DefSet}, we give
the basic definitions      and    settings in our   analysis.       In
Sec.~\ref{sec:numerical}, we  numerically  analyze  the  properties of
entanglement entropy  in      the QDM on      the  triangular lattice.
Especially,  we    examine  the  two   constructions  of   topological
entanglement entropy proposed by KP  and LW.  Furthermore, we consider
the entanglement entropy on a 
particular  topologically non-trivial area 
and   design another procedure to extract  $\gamma$, 
which, for QDMs,  turns  out to give an accurate value 
even in relatively small systems.
We then conclude in Sec. \ref{sec:conclusion}.

\section{Definitions and Settings}\label{sec:DefSet}

\subsection{Model}\label{sec:model}
We consider the QDM on the triangular lattice defined by the Hamiltonian:\cite{Rokhsar88,Moessner01}
\begin{align}\label{eq:QDM-tri}
H=\sum_{\mathrm{rhombi}} 
 \bigg[ &-t\left(\bigg|\QDMts\bigg\rangle\bigg\langle\QDMtz\bigg|+\mathrm{h.c.}\right) \notag \\ 
        &+v\left(\bigg|\QDMtz\bigg\rangle\bigg\langle\QDMtz\bigg|
                +\bigg|\QDMts\bigg\rangle\bigg\langle\QDMts\bigg|\right)
 \bigg],
\end{align}
where the sum runs over all rhombi consisting of two neighbouring triangles and we set $t>0$.
At the Rokhsar-Kivelson (RK) point $v=t$, a ground-state is given exactly 
by the equal-amplitude superposition of all the dimer coverings:\cite{Rokhsar88}
\begin{equation}\label{eq:RK-fn}
 \ket{\RK} \equiv \frac{1}{\sqrt{| \Ecal |}} \sum_{C\in\Ecal} \ket{C}, 
\end{equation}
where  $\Ecal$ denotes the set of  all the dimer coverings.  This wave
function       exhibits     exponentially-decaying         dimer-dimer
correlations\cite{Moessner01,Ioselevich02,Fendley02} and is    an example of liquid  with  no
broken symmetries.

This wave function is not the unique ground  state if the lattice has a
non-trivial  topology  (cylinder, torus, etc.).  Let   us focus on the
case of the torus   hereafter.    We draw two   incontractible   loops
$\Delta_1$ and  $\Delta_2$ which pass  through the  bonds and wind
around
the  torus   in  $x$    and   $y$   directions  respectively   as   in
Fig.~\ref{fig:tri_lattice}.   We   classify  $\Ecal$ into   four  sets
$\Ecal^p$ with $p=++,+-,-+,--$, depending on the  parity of the number
of   dimers crossing $\Delta_1$  and $\Delta_2$.    The resultant sets
$\Ecal^p$,  called  {\em topological  sectors},  are  not mixed by any
local dimer  move (and thus   by any term in  the  Hamiltonian).   The
spectrum of the Hamiltonian can  therefore be determined separately in
each sector.  At  the  RK point, the   ground-state in  each sector is
given by
\beqn\label{eq:RK-fn-sec}
 \ket{\RK;p} \equiv \frac{1}{\sqrt{| \Ecal^p |}} \sum_{C\in\Ecal^p} \ket{C}.
\eeqn
All  these  states      have    zero energy for     the    Hamiltonian
\eqref{eq:QDM-tri} and span  a four-dimensional ground-state manifold.
It has been shown analytically and  numerically that the degeneracy of
the ground  states and   the exponential   decay  of  the  dimer-dimer
correlation at the RK point persist in a finite range in the parameter
space, forming a liquid phase with gapped excitations
in                         $0.82(3)\lesssim                         v/t\le
1$.\cite{Moessner01,Ioffe02_Nature,Ralko05-06,Ioselevich02} Decreasing
$v/t$ further, the model  enters a  valence  bond crystal (VBC)  phase
with a large unit cell  (12 sites),
called  $\sqrt{12}\times\sqrt{12}$ VBC.\cite{Moessner01,Ralko05-06}

The  ground-state degeneracy in  the liquid phase  indicates that this
phase is topologically ordered.    It is  indeed  a
realization of  the     deconfined  phase  of   a   $\mathbb{Z}_2$ (Ising) gauge
theory,\cite{Wegner71} where the requirement that physical states must
be invariant    under gauge transformations is    played  by the dimer
hard-core constraint  and where the role of  the gauge flux piercing a
plaquette  is      played  by a  dimer-move      operator  around this
plaquette.\cite{Moessner02,Misguich02}   The      four    ground-states
correspond to the four possible choices to put (or  not to put) a vortex
through the two holes of the torus.

\subsection{Lattice}\label{sec:lattice}
\begin{figure}[h]
\centerline{
\includegraphics[width=\linewidth]{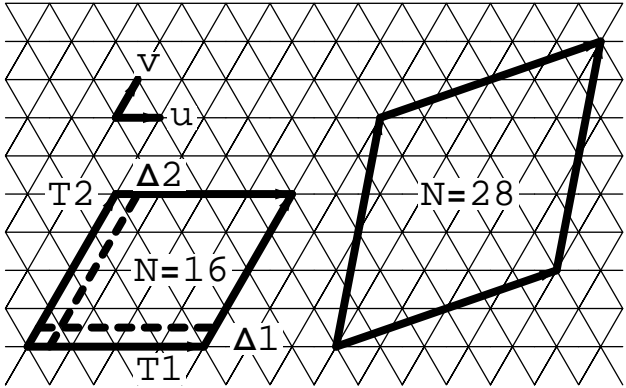}
}
\caption{\label{fig:tri_lattice}
Triangular lattices with periodic boundary conditions.
}
\end{figure}

The lattice is put on a torus and is defined by two vectors $\bm{T}_1$
and  $\bm{T}_2$  specifying the periodicity.    We mostly use lattices
which are symmetric under $120^\circ$ rotation, by setting
\begin{equation}\label{eq:vec-T}
 \bm{T}_1=l\bm{u}+m\bm{v},\quad
 \bm{T}_2=-m\bm{u}+(l+m)\bm{v},
\end{equation}
where $l$   and $m$ are  integers  and $\bm{u}$ and  $\bm{v}$ are unit
vectors as  shown in Fig.  \ref{fig:tri_lattice}.  The total number of
sites is    given by $N=l^2+lm+m^2$.   The   lattices we consider have
$N=16,28,36,48,52,64$,  which correspond respectively        to
$(l,m)=(4,0),(4,2),(6,0),(4,4),(6,2),(8,0)$.
In  Sec.~\ref{sec:nonlocal}, $N=100$ (corresponding to $(10,0)$) is also studied.

\subsection{Reduced density matrix}\label{sec:RDM}
To define a reduced density matrix (RDM) for  the QDM, we must specify
the local degrees of freedom of the model.  To this  end, we assign an
Ising  variable $\sigma_k$  to each  bond  $k$  of  the  lattice as in
Ref.~\onlinecite{Moessner02}  and identify  the  presence/absence of a
dimer  on  the  bond  as  $\sigma_k=+1$  and  $-1$, respectively.  Any
physical  configuration $\{\sigma_k\}$   must satisfy  the   hard-core
constraints: for  each site of the  lattice, there must be exactly one
bond with $\sigma_k=1$ emanating from it.  An area $\Omega$ is defined
as a set of {\it bonds}. We define the matrix element  of the RDM of a
ground-state $\ket{\Psi}$ as
\begin{equation}
 \bra{c_1}\rho_\Omega\ket{c_2} 
 = \sum_{\cbar} \bracket{c_1,\cbar}{\Psi} \bracket{\Psi}{c_2,\cbar}, 
\end{equation}
where $c_1$ and $c_2$ are dimer configurations on $\Omega$ and the sum
is over  all the dimer  configurations  $\cbar$ on  $\Omegabar$.  Note
that   we set  $\bracket{c,\cbar}{\Psi}=0$     if $(c,\cbar)$   is  an
unphysical configuration (violating the hard-core constraint).

Since the liquid  phase under consideration exhibits degenerate ground
states,  we must specify for which  state in the ground-state manifold
we calculate the entanglement entropy.   However, as long as the  area
is local, it was numerically demonstrated that  the RDMs are identical
for all  states in the ground-state  manifold, up to  a correction
which decays exponentially with the system size.\cite{Furukawa06} Thus
in this case we  can take any state  in the ground-state manifold.  At
the RK point,  which  we mainly consider  in the  following, we simply
take the ``equal-amplitude'' state  \eqref{eq:RK-fn}.  The RDM  of the
``equal-amplitude'' state can be calculated in a  way described in the
Appendix \ref{App:RDM-RK}, either by direct enumeration, or using Pfaffians.

\section{Numerical results}\label{sec:numerical}

Here we present our numerical results.
The idea of KP and LW should apply to the dimer liquid phase in $0.82(3)\lesssim v/t \le 1$. 
The topological entropy for this phase is expected to be $\gamma=\ln 2\simeq 0.6931$, 
reflecting $\mathbb{Z}_2$ topological order.
We mainly consider the RK point $v/t=1$
with exact ground-states \eqref{eq:RK-fn} or \eqref{eq:RK-fn-sec}, 
and calculate the entanglement entropies using the methods in Appendix \ref{App:RDM-RK}.
For $v/t<1$, we perform Lanczos diagonalization of the Hamiltonian \eqref{eq:QDM-tri} 
for small systems (up to $N=36$), 
and calculate the entanglement entropies in the ground-state.

\subsection{Circular areas}\label{sec:circular}

\begin{figure}[h]
\centerline{
\includegraphics[width=0.5\linewidth]{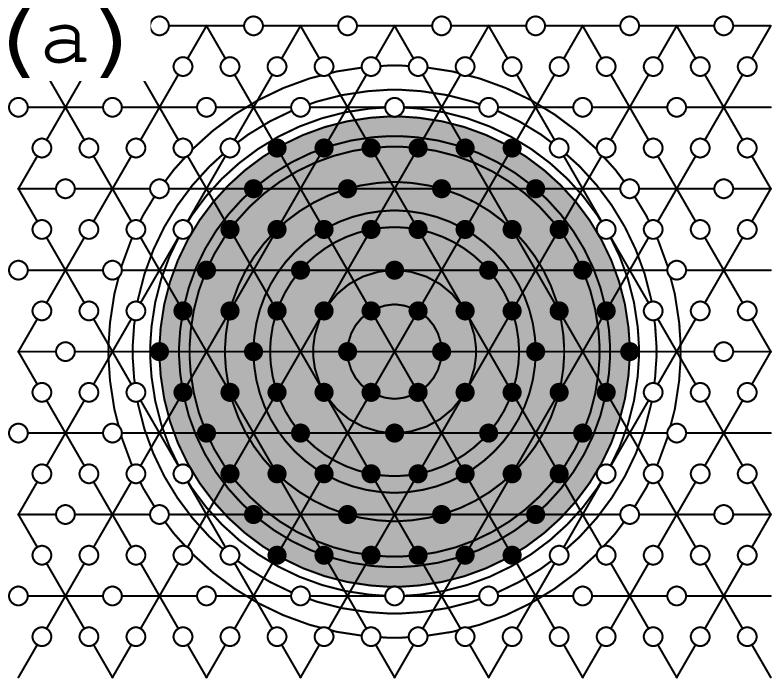}
\includegraphics[width=0.5\linewidth]{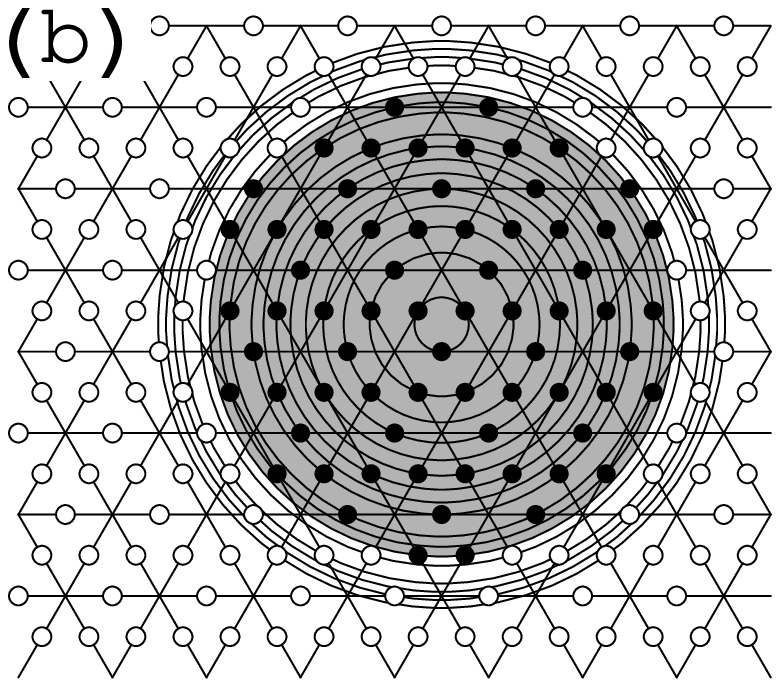}
}
\caption{\label{fig:circular-areas}
Circular  areas centered  at  (a) a  site  or  (b)  an  interior of  a
triangle.  As examples, areas with $R=2.5$ and $R=2.47$ are shaded for
(a) and (b), respectively.  }
\end{figure}

We first  consider the  entanglement entropy  on  disks (areas with no holes)  
and discuss how   the  entanglement entropy  scales  with  the extension   of the area.    
Calculations  were done   for the RK  wave function \eqref{eq:RK-fn}.   
As  the choice of  the area  $\Omega$, we define circular areas in the following way: 
we  draw a circle with a radius $R$ centred at a site or at an interior of a  triangle 
and regard every bond whose midpoint   is in   the  circle  as  an    element  of the    area; 
see Fig.~\ref{fig:circular-areas}.   
This definition causes an unavoidable ambiguity in the radius $R$ --- 
different radii can result in the same area. 
For example, the possible radius for the smallest site-centered area 
(consisting of six bonds) ranges in
$R_{\rm min}=0.5<R<\frac{\sqrt{3}}2=R_{\rm max}$.
Here we analyze the data taking this ambiguity into account.

\begin{figure}[h]
\centerline{\includegraphics[width=\linewidth]{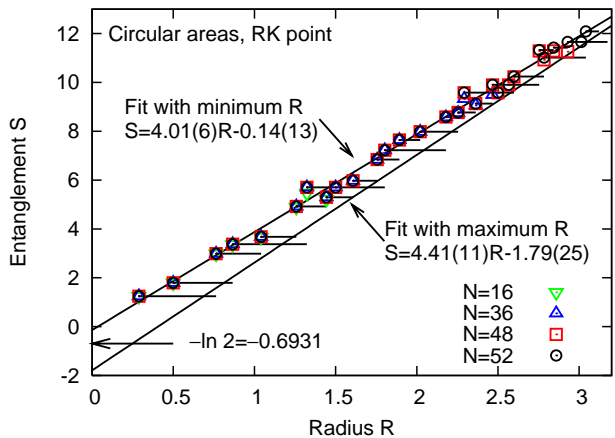}}
\caption{\label{fig:ent_circular}(color online)
Entanglement entropy on circular areas with radii $R$ at the RK point.
The ambiguity in $R$ is indicated by horizontal bars (only for $N=52$). 
The data for $N=52$ are fitted by lines using minimum or maximum radii. 
The resultant linear functions shown in the figure 
contains some numbers enclosed in parentheses, 
which indicates the standard errors in the last displayed digits.
}
\end{figure}

In Fig. \ref{fig:ent_circular}, the values of $S_\Omega$ on circular areas 
are plotted versus the radius $R$. 
The different symbols correspond to different system sizes (from $N=16$ to 52)
and an horizontal  bar specifies the interval $[R_{\rm min},R_{\rm max}]$
($N=52$ data points only, for clarity)
The data from  different system  sizes almost coincide,  
showing the smallness    of  the finite-size  effects. 
We fit the data for $N=52$ by a linear relation
using $R_{\rm min}$ or $R_{\rm max}$
We observe a rough agreement with the linear fitting in both cases 
as expected from the scaling form \eqref{eq:ent_scaling}. 
The lines intersect the vertical axis 
around $-0.1$ and $-1.8$ 
when using $R_{\rm min}$ and $R_{\rm max}$, respectively. 
These values sandwich the expected value $-\ln 2\simeq -0.6931$ but are both away from it.
We also fitted the data separately for site-centered and triangle-centered cases 
(not shown in the figure), 
but no essential difference was observed. 
These results show that a direct check of the scaling 
\eqref{eq:ent_scaling} is difficult.

In general, on a lattice, the boundary of $\Omega$  is made  of segments.  
If the sum of the segments is long enough, they contribute to the entanglement entropy 
by an amount proportional to the length. 
But in addition, we have to take into account the contribution coming from local correlations 
(between  the  regions $\Omega$  and $\Omegabar$) 
taking  place in the vicinity of the angles between successive segments. 
If $\Omega$ is large, the contribution from these angles may be small 
(of order $\mathcal{O}(L^0)$, compared to the boundary length $L$), 
but  this contribution will still be of the  same order as the topological term 
we are looking for.

In the present case (circular  areas), this  ambiguity in defining a
boundary length   on the  lattice  appears   as an ambiguity    in the
definition of $R$.
To compute $\gamma$  in a well-defined way,  
we need to turn to the constructions using plural areas, 
which we discuss in the next subsection.

\subsection{Construction of the topological entropy using plural areas}\label{sec:topological}
KP and  LW  proposed two  ways to  extract   the topological  constant
$\gamma$   independently    of    the   definition    of  the   boundary
length.\cite{Kitaev06,Levin06} The   idea is to  evaluate  $\gamma$ by
forming   an   appropriate linear    combination  of the  entanglement
entropies  of   different areas,  so that  the  boundary contributions
cancel out.

\subsubsection{Kitaev-Preskill construction}

In the KP construction,\cite{Kitaev06} we consider a circle 
and divide it into   three ``fans'',  $A$, $B$,  and   $C$.  
Then   we  form a  linear combination
\begin{equation}\label{eq:ent_topo_KP}
 S_\topo^\KP = S_A+S_B+S_C-S_{AB}-S_{BC}-S_{CA}+S_{ABC},
\end{equation}
where $S_{XY\cdots}$ denotes the  entanglement entropy on a  composite area $X\cup Y\cup\cdots$.    
In this combination, all the boundary contributions cancel out 
and a topological term $-\gamma$ should remain.
For example, let us consider the line separating $A$ and $B$.
The boundary contributions along this line appears in $S_A$ and $S_B$ with a plus sign 
and in $-S_{BC}$ and $-S_{CA}$ with a minus sign.
Some attention should be paid to the triple point,
in the vicinity of which the areas have different shapes
and thus possibly different local contributions.
Three areas form a 120 degree angle: $A$, $B$ and $C$; 
three areas form a 240-degrees angle: $BC$, $AC$ and $AB$.
However, recalling that the entanglement entropies of an area and its complement are the same, 
the entropy of $BC$ is equal to that of the complement of $BC$, 
which has the same shape with $A$ in the vicinity of the triple point. 
Thus the {\it local} contributions from $A$ and $BC$ in the vicinity of the triple point
should match.
The same argument applies to every line and every corner,
giving a cancellation of all the boundary contributions in Eqn. \eqref{eq:ent_topo_KP}. 
Assuming the scaling \eqref{eq:ent_scaling}, we expect $S_\topo^\KP=-\gamma$
(for a large enough radius).

\begin{figure}
\centerline{
\includegraphics[width=0.5\linewidth]{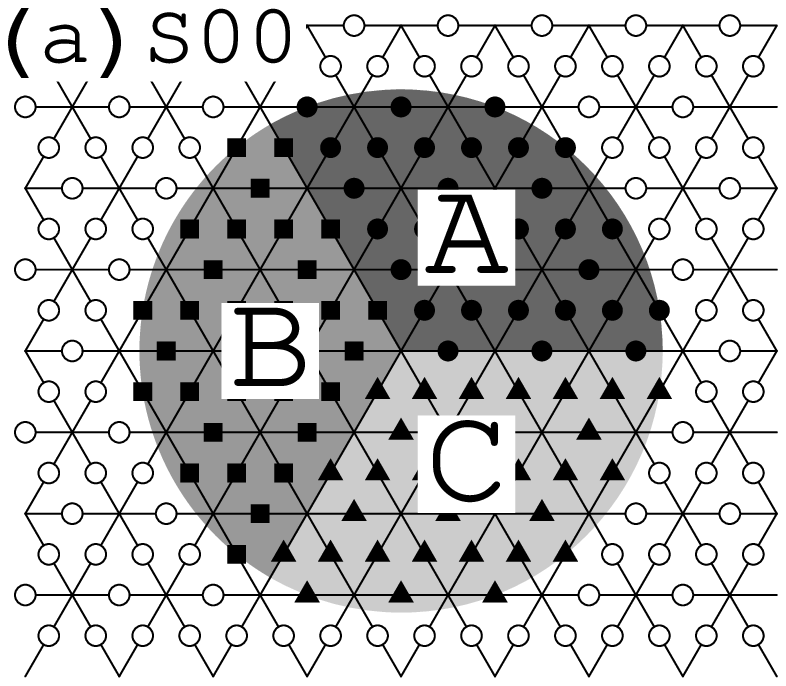}
\includegraphics[width=0.5\linewidth]{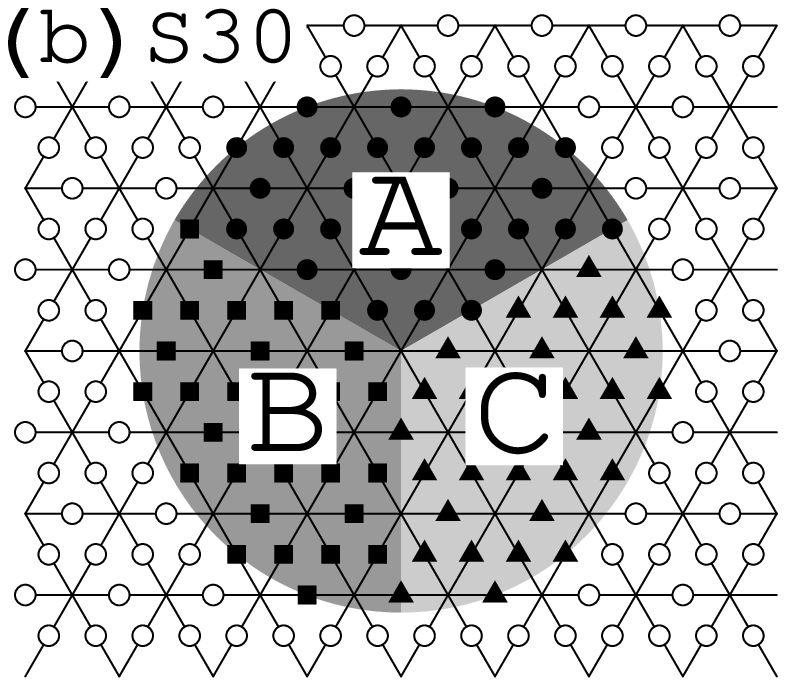}
}
\centerline{
\includegraphics[width=0.5\linewidth]{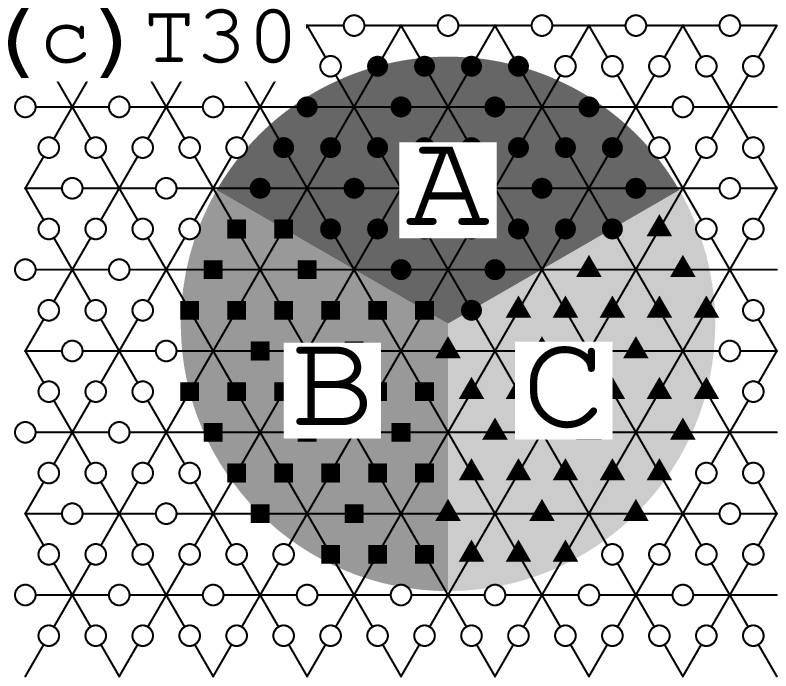}
\includegraphics[width=0.5\linewidth]{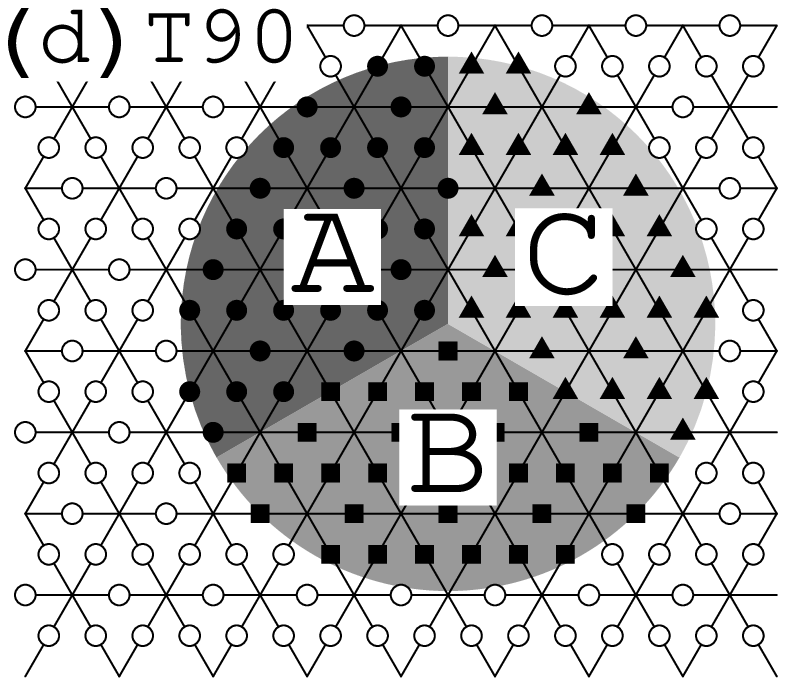}
}
\caption{Divisions of circular areas for the Kitaev-Preskill construction.
(a) and (b): site-centered, $R=2.78$.
(c) and (d): triangle-centered, $R=2.84$.}
\label{fig:circle_divide_KP}
\end{figure}

We apply this idea to the present model. 
We divide a circle by three lines emanating from the center as in Fig.~\ref{fig:circle_divide_KP}. 
These lines are placed at angles 
$\theta_0-0$,  $\theta_0+120^\circ-0$ and    $\theta_0+240^\circ-0$ 
measured from the (reference)  $\bold{u}$ direction.
Here ``$-0$'' represents an infinitesimal shift for avoiding collisions 
between the points (midpoints of bonds) and the boundaries.
For example, points at an angle $\theta_0$ belong to $A$, not to $C$.
We  take  $\theta_0=0^\circ$   or $30^\circ$ for site-centered circles 
(referred to as  ``S00''   and ``S30'') 
and $\theta_0=30^\circ$  or  $90^\circ$ for triangle-centered circles 
(``T30'' and ``T90'').  
In  these  settings, the parts $A,B,C$ are equivalent under  $120^\circ$ rotation, 
and  we thus only  need to calculate  $S_\topo^\mathrm{KP}=3S_A-3S_{AB}+S_{ABC}$.    

\begin{figure}
\centerline{\includegraphics[width=\linewidth]{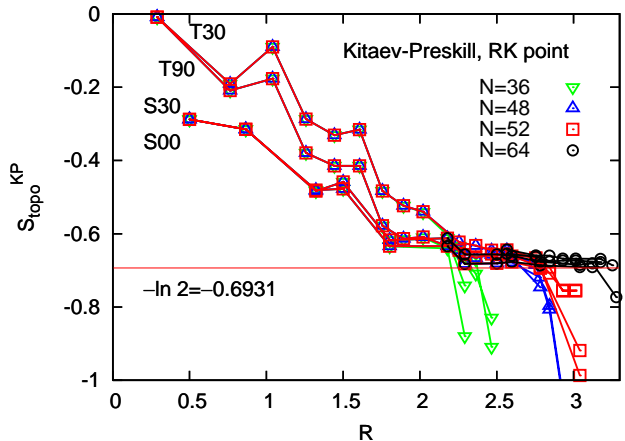}}
\caption{(color online) 
Topological entanglement entropy from the Kitaev-Preskill construction \eqref{eq:ent_topo_KP} 
at the RK point. 
Examples of areas are shown in Fig. \ref{fig:circle_divide_KP}.
Some explicit values for large radii are shown in Table \ref{tab:ent_topo_KP}.}
\label{fig:ent_topo_KP}
\end{figure}

\begin{table}
\caption{\label{tab:ent_topo_KP}
Some values of $-S_\topo^\KP$ at the RK point, 
divided by the expected value $\ln 2$.
The data for large radii are shown, 
and excellent agreement with the expectation can be seen.
}
\begin{minipage}{0.47\linewidth}
\begin{ruledtabular}
\begin{tabular}{ccc}
\multicolumn{3}{c}{S00 case} \\
Radius $R$ & \multicolumn{2}{c}{$-S_\topo^\KP/\ln 2$}\\
 & $N=52$ & $N=64$ \\
 \hline
2.18 & 0.9143 & 0.9143\\
2.29 & 0.9839 & 0.9835\\
2.50 & 0.9822 & 0.9822\\
2.60 & 0.9765 & 0.9760\\
2.78 & 1.0014 & 0.9897\\
3.04 & 1.3252 & 0.9967\\
3.12 &        & 0.9967
\end{tabular}
\end{ruledtabular}
\end{minipage}
\begin{minipage}{0.47\linewidth}
\begin{ruledtabular}
\begin{tabular}{ccc}
\multicolumn{3}{c}{T30 case} \\
Radius $R$ & \multicolumn{2}{c}{$-S_\topo^\KP/\ln 2$} \\ 
 & $N=52$ & $N=64$ \\
 \hline
2.57 & 0.9291 & 0.9283\\
2.75 & 0.9618 & 0.9513\\
2.84 & 0.9965 & 0.9518\\
2.93 & 1.0910 & 0.9635\\
3.01 & 1.0910 & 0.9635\\
3.18 & & 0.9649\\
3.25 & & 0.9898
\end{tabular}
\end{ruledtabular}
\end{minipage}
\end{table}

\begin{figure}
\centerline{\includegraphics[width=\linewidth]{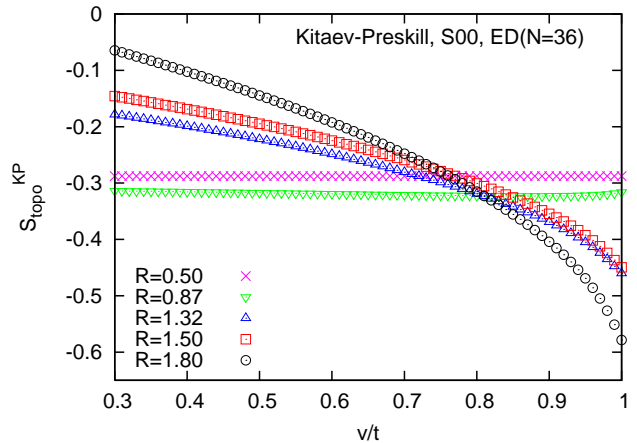}}
\centerline{\includegraphics[width=\linewidth]{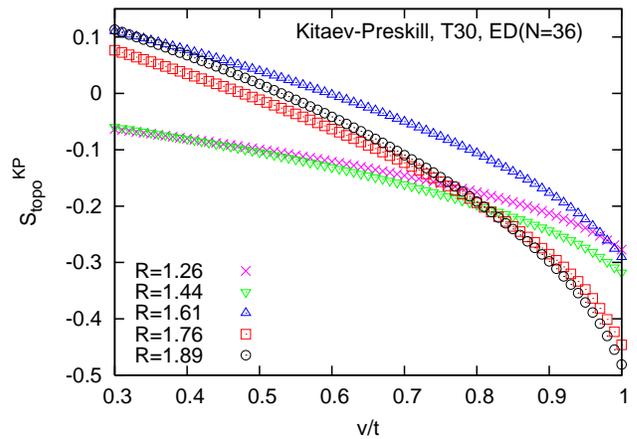}}
\caption{(color online) 
Kitaev-Preskill topological entropy \eqref{eq:ent_topo_KP} as a function of $v/t$ for $N=36$. 
In the large-$R$ limit, $S_\topo^\KP$ is expected jump from $\-\ln 2$ 
in $\mathbb{Z}_2$ liquid phase $0.82(3)\lesssim v/t \le 1$,
to some positive value in the VBC phase $v/t\lesssim 0.82(3)$.}
\label{fig:ent_topo_KP_ED}
\end{figure}

We first consider the case of the RK wave function \eqref{eq:RK-fn}.
In Fig. \ref{fig:ent_topo_KP}, the data of $S_\topo^\KP$ are plotted versus the radii $R$ of the circles.
As in the case of circular areas presented in Fig. \ref{fig:ent_circular}, 
finite-size effects are very small --
except for the case where the circle $ABC$ occupies a substantial part of the system, 
the data from different $N$'s almost coincide. 
In the largest system $N=64$, we can regard the data up to $R\lesssim 3.1$ 
as good approximation to the values in the infinite system.
In all the cases, $S_\topo^\KP$ decreases almost monotonically with $R$ 
and for large radii (specifically, $2.2\lesssim R\lesssim 3.1$) shows values which are very close to $-\ln 2$ 
(see Table \ref{tab:ent_topo_KP}), 
the expected value for a $\mathbb{Z}_2$ topologically ordered state.

Next we consider the region $v/t<1$ of the Hamiltonian \eqref{eq:QDM-tri}. 
In $\mathbb{Z}_2$ liquid phase $0.82(3)\lesssim v/t \le 1$, 
$S_\topo^\KP$ is expected to show $-\ln 2$ in the large-$R$ limit.
On the other hand, in $\sqrt{12}\times\sqrt{12}$ VBC phase $v/t\lesssim 0.82(3)$, 
where discrete symmetries are spontaneously broken, 
the finite-size ground-state can be approximated by 
a linear superposition of 12-fold symmetry-broken states.
In such a state, we conjecture that the entanglement entropy on a disk $\Omega$ 
scales as $S_\Omega \simeq \alpha L + \ln d$ in the large-area limit, 
where $d$ is the ground-state degeneracy and is equal to 12 in the present case.
The constant term $\ln d$ is {\em not topological} 
in the sense that   the same  value would appear even if $\Omega$  had  another geometry, 
unlike  $-\gamma$  in  Eqn.~\eqref{eq:ent_scaling}.   
Note also that this  constant  is  positive, in contrast to  the negative topological term $-\gamma$. 
Assuming this,  the combination \eqref{eq:ent_topo_KP} should give $\ln d$ in a symmetry-broken phase.
Thus, $S_\topo^\KP$ is expected to jump from a negative (topological) value $-\ln 2$ 
to a positive (non-topological) value $\ln 12$
along with the transition from the liquid phase to the VBC phase.
We performed Lanczos diagonalization of the Hamiltonian \eqref{eq:QDM-tri} 
for a lattice with $N=36$ 
(which is the maximum size in our exact diagonalization calculation 
and is compatible with $\sqrt{12}\times\sqrt{12}$ VBC ordering), 
and calculated $S_\topo^\KP$ in the ground-state, 
which lies in the sector $p=--$ in both the VBC and liquid phases on this lattice.
The results are shown for two types of areas (``S00'' and ``T30'') in Fig. \ref{fig:ent_topo_KP_ED}.
Because the system and area sizes are rather small, 
we do not observe a jump at the transition. 
However, we can already observe some tendency: for fixed $v/t$, 
$S_\topo^\KP$ tends to decrease as a function of $R$ in the liquid side 
while it tends to increase in the VBC side.
Some positive values of $S_\topo^\KP$ in the VBC phase are also seen in ``T30'' case.

\subsubsection{Levin-Wen construction}

\begin{figure}
\centerline{
\includegraphics[width=0.5\linewidth]{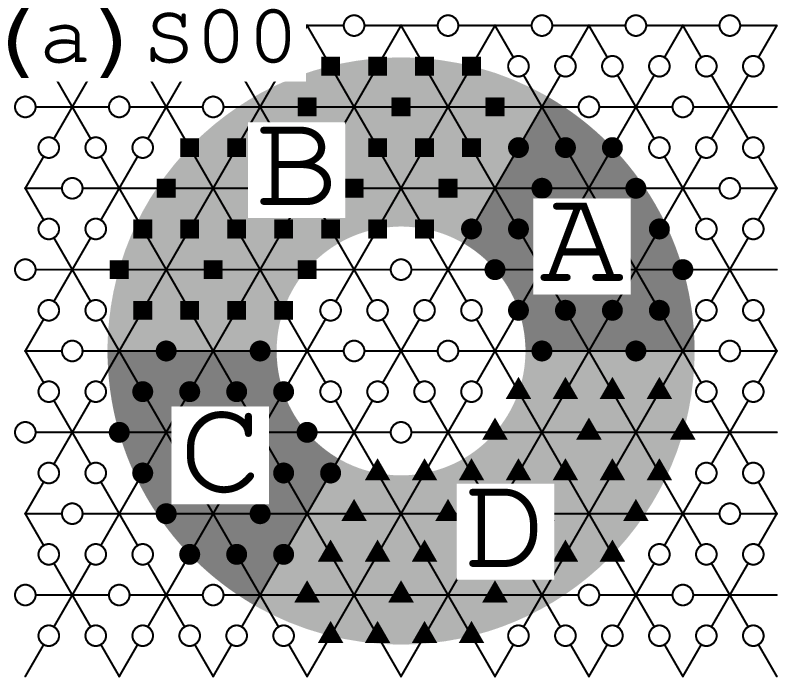}
\includegraphics[width=0.5\linewidth]{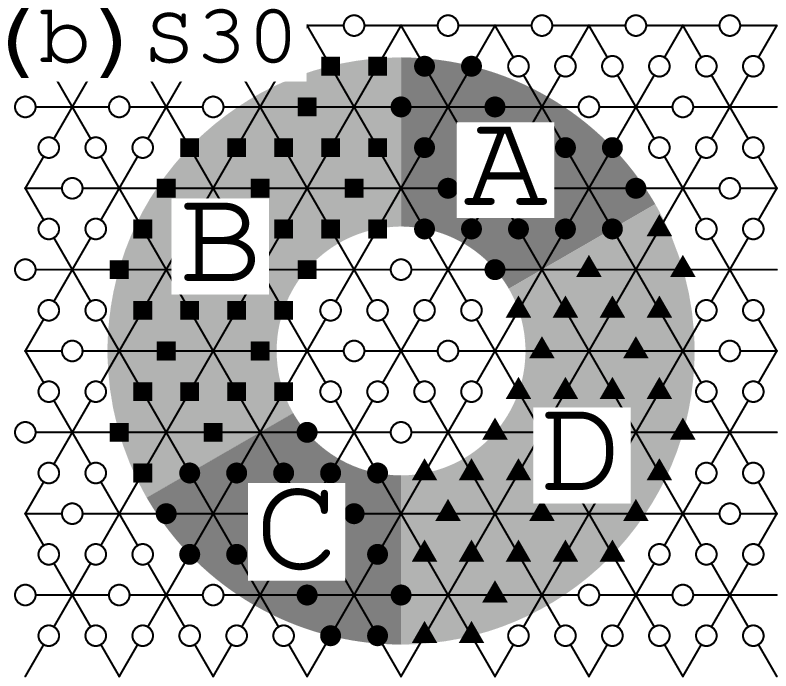}
}
\caption{Division of annular areas for the Levin-Wen construction.}
\label{fig:annular_divide_LW}
\end{figure}

In the LW construction \cite{Levin06}, we  consider an annulus divided
into four pieces as in Fig.  \ref{fig:annular_divide_LW}, and form a
combination
\begin{equation}\label{eq:ent_topo_LW}
 S_\topo^\LW=S_{ABCD}-S_{ABC}-S_{CDA}+S_{AC}.
\end{equation}
This combination    is   guaranteed  to   be   non-positive   from the
strong subadditivity inequality of entanglement entropies,\cite{Nielsen00} namely, 
\begin{equation}
 S_\topo^\LW=S_{X\cup Y}-S_X-S_Y+S_{X\cap Y}\le 0, 
\end{equation}
where $X=A\cup  B\cup   C$ and  $Y=C\cup  D\cup   A$.  
The combination \eqref{eq:ent_topo_LW}  is  expected  to    give  $-2\gamma$  for    a
topological phase and zero for a conventional phase
(disordered, or with some symmetry-breaking order).

\begin{figure}
\centerline{\includegraphics[width=\linewidth]{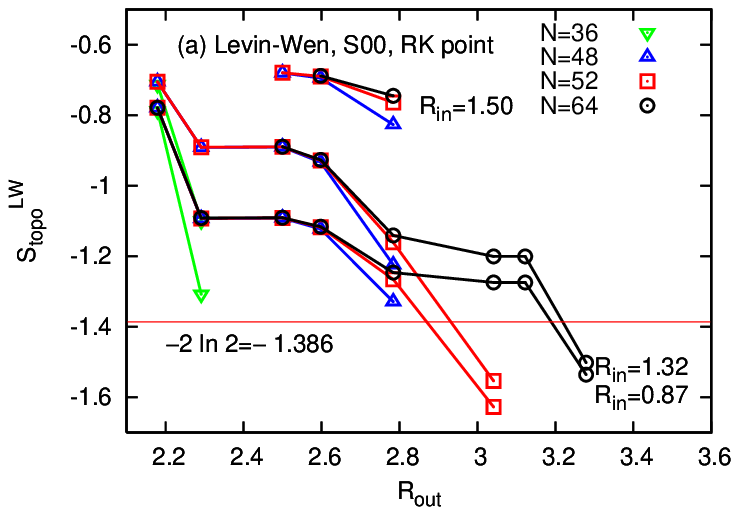}}
\centerline{\includegraphics[width=\linewidth]{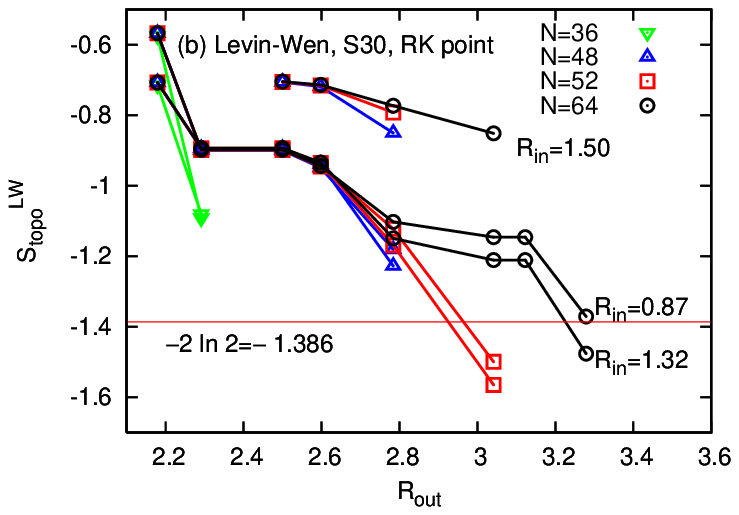}}
\caption{Topological entanglement entropy from the Levin-Wen construction.}
\label{fig:ent_topo_LW}
\end{figure}

In Fig.   \ref{fig:annular_divide_LW}, an annulus   is divided by four lines at angles
$\theta_0-0,\theta_0+60^\circ+0,\theta_0+180^\circ-0,\theta_0+240^\circ+0$. 
We consider only site-centered annuli,  and we set $\theta_0=0^\circ$  or $30^\circ$ 
(again referred to as ``S00''  and ``S30'').  
The result for the RK wave function is shown in Fig.~\ref{fig:ent_topo_LW}.  
$\Rin$ and $\Rout$ denote the inner and outer radii of the annulus respectively, 
and $S_\topo^\LW$ 's are plotted as a function of $\Rout$. 
Up to $\Rout\lesssim 3.1$, where the data for $N=64$ well approximate the values in the infinite system, 
we observe that $S_\topo^\LW$  monotonically decreases with $\Rout$  and approaches $-2\ln 2$.  
Unfortunately, the convergence to $-2\ln 2$ is not very clear up to this radius. 
In the LW construction, the requirement for the convergence is $\xi << \Rin, \Rout-\Rin, L-2\Rout$, 
where $\xi$ is the correlation length ($\simeq 1$ at RK point)
and $L=\sqrt{N}$ is the linear system size (or equivalently, the maximum possible $2\Rout$).
Thus, the LW construction suffers from stronger finite-{\em area} (not finite-$N$) effects than 
the KP construction which just requires $\xi << R, L-2R$.


\subsection{Zigzag area}\label{sec:nonlocal}

\begin{figure}
\centerline{
\includegraphics[width=0.8\linewidth]{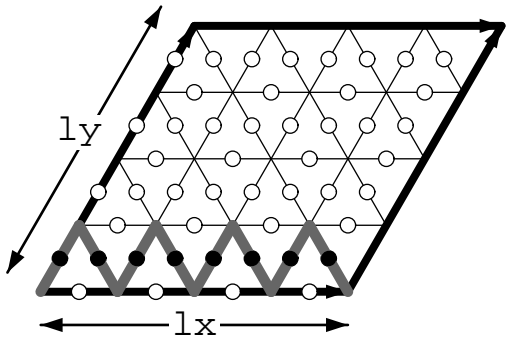}
}
\caption{\label{fig:zigzag_area}
Zigzag area on a lattice with $T_1=l_x\bold{u}$ and $T_2=l_y\bold{v}$.
}
\end{figure}

\begin{figure}
\centerline{\includegraphics[width=\linewidth]{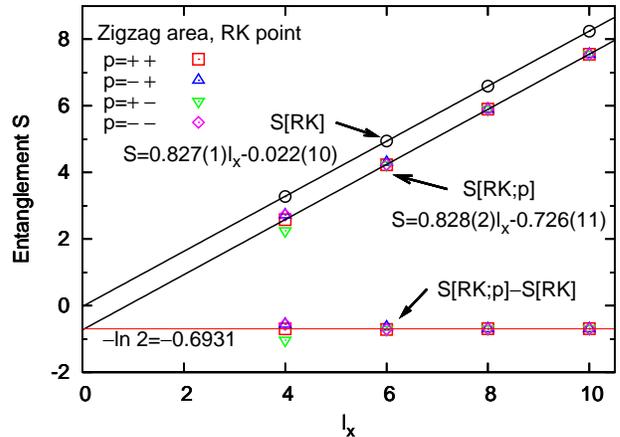}}
\caption{\label{fig:ent_RK_nonloc}
(color online)
Entanglement entropies on a zigzag area at the RK point.
The upper line is a linear fit  to $S[\RK]$. The lower  one is a fit to
$S[\RK;p=++]$ when $l_x=l_y$ is  multiple  of 4 and  $S[\RK;p=--]$
otherwise.  
The topological constant estimated from the latter fit is 
$\gamma=0.73\pm 0.01$.
This  particular choice of $p$  as a  function of $l_x$ is
motivated  by  the fact   that, when  $v/t<1$,  it corresponds  to the
ground-state sector.
Explicit values of $S[\RK]-S[\RK;p]$ are shown in Table \ref{tab:ent_RK_nonloc}.
%
}
\end{figure}

We design a different way to evaluate $\gamma$ using a thin ``zigzag'' area $\Omega$
winding around the torus as in Fig.~\ref{fig:zigzag_area}. 
This area is  invariant by translation  in  the $x$ direction and  all
points   (black  circles    in  Fig.~\ref{fig:zigzag_area}) are   {\it
equivalent    by symmetry}.  In contrast  to the    more complicated areas
considered   before,     we expect   the   boundary   (i.e., 
non-topological)   contribution to $S_\Omega$   to  be precisely proportional to
$l_x$, when  $l_x$ is sufficiently larger  than the correlation length
$\xi$.  In this new  geometry, the thermodynamic behavior is  obtained
as soon as $\xi\ll  l_x,l_y$, 
whereas the KP construction requires $\xi << R, L-2R$,  
which is difficult to reach in exact diagonalization up to $N=36$.
Since the area is topologically non-trivial (it contains
the incontractible cut $\Delta_1$), the value of $S_\Omega$ depends on
the      choice    of   the   ground       state,   even   for   large
systems.  We  calculate  the  entanglement entropies on
this area  in  the  ground-states $\ket{\RK}$  and  $\ket{\RK  ;p}$ on
isotropic    lattices $l_x=l_y$, and    write  them  as  $S[\RK]$  and
$S[\RK;p]$    respectively.    The      results    are   plotted    in
Fig.~\ref{fig:ent_RK_nonloc}.   As anticipated, $S_\Omega$ appears  to
be  almost perfectly  linear  in $l_x$ (compared  with the  results of
Fig.~\ref{fig:ent_circular}).    Moreover,   we observe    that    the
topological constant $\gamma$ can  be extracted in two different ways:
a)  by extrapolating (through a  linear fit) $S[\RK;p]$ at ``$l_x=0$''
or b)  by $-\gamma\simeq S[\RK;p]-S[\RK]$. These  two  follow from the
scaling forms
\begin{equation}\label{eq:ent_zigzag}
\begin{split}
 S=\alpha_1 l_x ~~ &\mathrm{for}~~ \ket{\RK},\\
 S=\alpha_1 l_x-\gamma ~~ &\mathrm{for}~~ \ket{\RK ;p},
\end{split}
\end{equation}
where   $\alpha_1$  is a  non-universal constant.   A   similar scaling was obtained
rigorously by  Hamma {\it et al.} \cite{Hamma05}  for a ``ladder'' area
in Kitaev's model on the  square lattice.  Here  we confirmed that  it
holds  accurately  even in a   system with a {\it  finite} correlation
length.  The  scaling forms~\eqref{eq:ent_zigzag}  provide an accurate
way to calculate the topological constant  $\gamma$ even in relatively
small  systems.  The condition  (satisfied by QDM) is that topological
sectors must be well defined and not mixed by the Hamiltonian, so that
one can label the ground-states by their sectors.
Computing $\gamma$ from  the  largest system ($l_x=l_y=10$)  gives our
best    estimate    of    the   topological     entanglement   entropy
$S[\RK]-S[\RK;p=--]=0.6939$ (to be compared with $\ln 2=0.6931$);
see Table \ref{tab:ent_RK_nonloc}.

\begin{table}
\caption{\label{tab:ent_RK_nonloc}
Values of $S[\RK]-S[\RK;p]$ on a zigzag area, divided by $\ln 2$. 
The sector $p$ used in Fig. \ref{fig:ent_RK_nonloc} 
is indicated by $*$, and gives the best estimate in most cases.
}
\begin{ruledtabular}
\begin{tabular}{rllll}
 & \multicolumn{4}{c}{$(S[\RK]-S[\RK;p])/\ln 2$} \\ 
\multicolumn{1}{c}{$l_x~(=l_y)$} & \multicolumn{1}{c}{$++$} & \multicolumn{1}{c}{$-+$} 
                                 & \multicolumn{1}{c}{$+-$} & \multicolumn{1}{c}{$--$} \\
 \hline
4  & 1.0024$^*$ & 0.8051 & 1.4910 & 0.8051\\
6  & 1.0315     & 0.9248 & 1.0315 & 1.0212$^*$\\
8  & 0.9944$^*$ & 1.0022 & 1.0017 & 1.0022\\
10 & 0.9981     & 1.0028 & 0.9981 & 1.0011$^*$
\end{tabular}
\end{ruledtabular}
\end{table}

\begin{figure}
\begin{center}
\includegraphics[width=\linewidth]{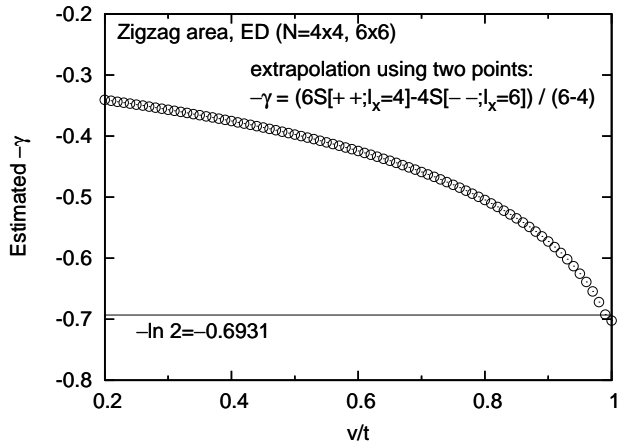}
\end{center}
\caption{\label{fig:ent_diag_nonloc}
The topological term $-\gamma$ as a function of $v/t$, 
estimated by using zigzag areas for $l_x=4$ and 6.
}
\end{figure}

As another application, we use the scaling forms \eqref{eq:ent_zigzag}
to evaluate the topological term $-\gamma$ 
in the region $v/t<1$.
We performed Lanczos diagonalization for lattices with $l_x=l_y=4$ and $l_x=l_y=6$.
For  $v/t<1$, the ground-state lies  
in the sector $p=++$ for $l_x=4$  and $p=--$ for $l_x=6$.  
We therefore  compute the entropies  $S[p=++;l_x=4]$  and  $S[p=--;l_x=6]$ on the zigzag areas 
and approximate the topological term $-\gamma$ 
by a linear extrapolation to ``$l_x=0$''.  
In the thermodynamic limit, the constant term extracted in this way 
is expected to jump from $-\ln 2$ to a positive value, 
as in the case of Fig.~\ref{fig:ent_topo_KP_ED}.
However, the $\sqrt{12}\times\sqrt{12}$ VBC ordering is compatible only with lattices 
where $l_x=l_y$ is a multiple of 6, 
and a linear relation $S_\Omega\simeq \alpha_1 l_x +\ln d$ holds only for such lattices.
\footnote{
The zigzag area under consideration is not a disk 
(the  width is too small to  contain  one unit cell of  the $\sqrt{12}\times\sqrt{12}$ crystal)  
but  all the $d=12$  VBC patterns can be distinguished by some appropriate observable defined on this area.
We thus expect the same scaling as for disks.
}
The lattice with $l_x=l_y=4$ is thus out of this scaling, 
and the present estimation of the constant term is invalid for the VBC phase.
Still, it can be  used in the liquid  phase.   The result is shown  in
Fig.~\ref{fig:ent_diag_nonloc}.  A value close  to $\ln2$ is recovered
at the RK point but  it decreases smoothly when  decreasing $v/t$.  No
clear signature of a  transition out of  the topological liquid can be
seen.   Larger  system  sizes are   probably  required  to  locate the
transition  with  this method. The problem  probably  lies in  a rapid
increase of the  dimer-dimer  correlation  length (and  thus  stronger
finite-size effects)  when  moving  away from   the  RK  point  in the
direction of the VBC phase.


\section{Summary and conclusions}\label{sec:conclusion}

The   concept  of   topological  entanglement  entropy   was  recently
introduced  by KP and   LW  as a  way   to  detect and characterize   
topological  order  from   a  ground-state  wave-function.   We   have
illustrated  numerically how this approach   works in the  case of the
$\mathbb{Z}_2$ liquid phase of the QDM on  the triangular lattice.  We
found that, due to lattice discretization, the topological entropy
$\gamma$  cannot be obtained  from a  direct fit  to the scaling  form
$S\simeq \alpha L -\gamma$.  Instead,  it is necessary to combine  the
entropies on plural areas to cancel  out the boundary contributions,
as suggested  by  KP and LW.    
In particular, for  the KP construction,  we clearly  observed that in
the large-area limit the topological entanglement entropy converges to
$-\ln   2$  expected for $\mathbb{Z}_2$   topological  order.  We also
proposed a  procedure to  evaluate  the  topological entropy  using  a
topologically non-trivial ``zigzag''   area, which gives   an accurate
value even in small systems.
For a system of linear size  $l_x=10$, the later method provided
an estimate of the topological entanglement entropy $0.6939$
at the RK point, 
in remarkable agreement with the expected value ($\ln 2=0.6931$).

In  addition to illustrating  the concept  of topological entanglement
entropy  in a  ``realistic'' model, the   present  analysis offers  an
evidence   of $\mathbb{Z}_2$  topological   order in   the QDM  on the
triangular  lattice   from  a   new  perspective.
Although   the   existence of
topological degeneracy,\cite{Moessner01}   the  analogy between   this
model and a $\mathbb{Z}_2$ gauge theory\cite{Moessner02}   and the absence of
{\it any} broken symmetry\cite{Furukawa06} were already known, the
present work confirms the $\mathbb{Z}_2$ structure in the ground-state
wave-function itself.

\begin{acknowledgments}
The authors are grateful  to C.~Lhuillier, M.~Oshikawa and V.~Pasquier
for  valuable  discussions from  the  early  stage  of  this  work and 
for critical reading of the manuscript.
SF thanks S. Ryu for useful discussions.
The authors also acknowledge insightful comments from anonymous referees, 
which helped the authors to reexamine the calculations 
and to improve the manuscript with refined results. 
Most  of this work was done  
while SF was  at CEA Saclay and at Universit\'e P. \&  M. Curie (Paris), 
under the support of  an exchange program, ``Coll\`ege Doctoral
Franco-Japonais'', and SF is thankful  for the kind hospitality  there.
SF was also supported  by a 21st  Century COE  Program at  Tokyo Tech,
``Nanometer-Scale Quantum Physics'' from the MEXT of Japan.
\end{acknowledgments}

\newcommand{\OS}{\Lambda}

\appendix
\section{Reduced Density Matrix of the RK Wave Function}\label{App:RDM-RK}

\begin{figure}
\centerline{
\includegraphics[width=\linewidth]{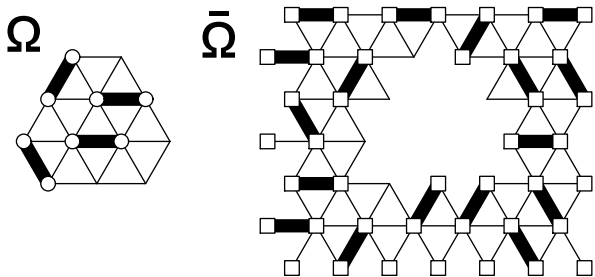}
}
\caption{\label{fig:occupied_sites}
Left (right): dimer configuration $c$ ($\cbar$) on $\Omega$ ($\Omegabar$) 
and their ``occupied sites'' $\OS (c)$ ($\OS (\cbar)$), marked with circles (squares).
In this picture, $\OS (c)$ and $\OS (\cbar)$ are compatible and thus $(c,\cbar)$ is physical.
}
\end{figure}

In this appendix, we derive a simple expression for  the RDM of the RK
wave function  \eqref{eq:RK-fn}, and describe two methods for calculating it.  
A  dimer configuration $C$   on the
entire system  can be divided  into the configurations on $\Omega$ and
$\Omegabar$:
\begin{equation}
 C\in\Ecal \to c\in\Ecal_\Omega, \cbar\in\Ecal_\Omegabar,
\end{equation}
where $\Ecal_\Omega$   ($\Ecal_\Omegabar$) denote the  set of  all the
possible dimer  configurations  on  $\Omega$  ($\Omegabar$).   Now  we
consider   the    inverse   mapping:  given   $c\in\Ecal_\Omega$   and
$\cbar\in\Ecal_\Omegabar$, under  what   condition  is $(c,\cbar)$   a
physical    configuration?  This condition   is   given  in terms   of
``occupied  sites'' of  dimer     configurations as follows.    For  a
configuration  $c\in\Ecal_\Omega$, we define $\OS  (c)$  as the set of all
the     sites  occupied     by    dimers  in   $c$,     as    shown in
Fig.~\ref{fig:occupied_sites}.  We similarly  define $\OS (\cbar)$ for
$\cbar\in\Ecal_\Omegabar$.  In order for  $(c,\cbar)$ to  be physical,
a) $\OS (c)$ and $\OS (\cbar)$ should not overlap with each other, and
b) the sum of $\OS  (c)$ and $\OS (\cbar)$ should  cover all the sites
of the lattice. Then we can rewrite the wave function \eqref{eq:RK-fn}
as
\begin{equation}
 \ket{\RK} = \frac1{\sqrt{| \Ecal |}}
 \sum_{\begin{matrix} {c\in\Ecal_\Omega, \cbar\in\Ecal_\Omegabar}\\{\OS (c)\sqcup\OS (\cbar)=X_s}\end{matrix}} \ket{c}\ket{\cbar},
\end{equation}
where  $X_s$  is the  set of all  the  sites.  If we  list  up all the
possible $\OS  (c)$ and write them as $\OS_i~(i=1,2,\cdots)$, we can
divide the summation as
\begin{equation}
 \ket{\RK} = \frac1{\sqrt{| \Ecal |}} \sum_i
 \sum_{\begin{matrix} c\in\Ecal_\Omega \\ \OS (c)=\OS_i \end{matrix}} \ket{c}
 \sum_{\begin{matrix} \cbar\in\Ecal_\Omegabar \\ \OS (\cbar)=X_s\setminus\OS_i \end{matrix}} \ket{\cbar}.
\end{equation}
We introduce
\begin{equation}\label{eq:local_RK_fn}
\begin{split}
 \Ecal_\Omega^i &\equiv \{ c\in\Ecal_\Omega | \OS (c)=\OS_i \}, \\
 \Ecal_\Omegabar^i &\equiv \{ \cbar\in\Ecal_\Omegabar | \OS (\cbar)=X_s\setminus\OS_i \},
\end{split}
\end{equation}
and we define normalized states on $\Omega$ and $\Omegabar$ as 
\begin{equation}
\begin{split}
 \ket{\psi_\Omega^i} \equiv \frac1{\sqrt{|\Ecal_\Omega^i|}} \sum_{c\in\Ecal_\Omega^i} \ket{c},~~
 \ket{\psi_\Omegabar^i} \equiv \frac1{\sqrt{|\Ecal_\Omegabar^i|}} \sum_{\cbar\in\Ecal_\Omegabar^i} \ket{c}.
\end{split}
\end{equation}
Then we arrive at the Schmidt decomposition~:
\begin{equation}
 \ket{\RK} = \sum_i \sqrt{\lambda_i} \ket{\psi_\Omega^i} \ket{\psi_\Omegabar^i},~~\mathrm{with}~~
 \lambda_i \equiv \frac{|\Ecal_\Omega^i| \cdot |\Ecal_\Omegabar^i|}{|\Ecal|}.
\end{equation}
The RDM of this state reads 
\begin{equation}
\begin{split}
 \rho_\Omega &= \Tr_\Omegabar \ket{\RK}\bra{\RK}
              = \sum_i \bracket{\psi_\Omegabar^i}{\RK} \bracket{\RK}{\psi_\Omegabar^i}\\
             &= \sum_i \ket{\psi_\Omega^i} \lambda_i \bra{\psi_\Omega^i}.
\end{split}            
\end{equation}
This expression   is already diagonal and   the  $\lambda_i$'s are the
eigenvalues of $\rho_\Omega$.  The entanglement  entropy is then given
by  $S_\Omega=-\sum_i\lambda_i\ln\lambda_i$.  Since $\lambda_i$'s  are
expressed  using the number  of  dimer coverings for   a given set  of
occupied sites, the task has been reduced to counting dimer coverings.
This can be done by direct enumeration using a  recursive algorithm or by
a Pfaffian  method.

The Pfaffian method uses  the fact that  the number of dimer coverings
is given by the
the Pfaffian of an   adjacency matrix with appropriate signs  (entries
are   $\pm 1$     if the      two   sites   are  connected,   and    0
otherwise).\cite{Kasteyleyn61}   Counting only configurations $(c,c')$
such that $\OS(c)=\OS_i$ and  $\OS(\cbar)=X\setminus\OS_i$ can be done
by  removing  some   bonds of   the   lattice  (setting to   zero  the
corresponding matrix element): if a site $x$ belongs to $\OS_i$, there
cannot be any dimer between $x$ and a site $y$ if $(xy)$ is not a bond
of  $\Omega$.  In the same  way,  any bond $(xy)\in\Omega$ involving a
site    $x\notin\OS_i$   must     be  switched    off.   The   product
$|\Ecal_\Omega^i|\cdot |\Ecal_\Omegabar^i|$ is  thus obtained from the
Pfaffian  of the modified  adjacency matrix  above.\footnote{Since the
lattices we consider have periodic boundary conditions, four Pfaffians
(corresponding  to periodic/antiperiodic boundary  conditions in  both
directions) must in fact be combined to get the number of coverings in
a given  topological sector.}  It is clear  that sites in the ``bulk''
of $\Omega$  are necessarily included   in all $\OS_i$ (otherwise  the
number of configurations is zero) and those in the bulk of $\Omegabar$
are necessarily excluded. There is a choice only  for the sites in the
vicinity of the boundary between $\Omega$ and $\Omegabar$ (sites which
are both connected to bonds $\in\Omega$ and bonds $\notin\Omega$). The
number of possible  $\OS_i$ therefore scales  as $2^{P(\Omega)}$ where
$P(\Omega)$ is the  ``perimeter'' of $\Omega$.  Since the  calculation
of each  Pfaffian requires of the  order of $\sim N^3$ operations (see
Ref.~\onlinecite{Maeda03} for an explicit algorithm) the computer time
required to obtain the RDM (and its spectrum) scales as $\sim N^3\cdot
2^{P(\Omega)}$.  This method is   thus appropriate to  study ``small''
areas      in      ``large''       systems.       The      results  of
Fig.~\ref{fig:ent_RK_nonloc}  for  zigzag    areas  with $l_x=8$   and
$l_x=10$ were obtained by this method.

The  direct enumeration algorithm   searches and counts physical dimer
configurations one  by  one for a given  set  of  occupied sites.  The
enumeration is done separately  for $\Omega$ and $\Omegabar$,  and the
required  time for each area is  almost proportional  to the number of
dimer coverings, $|\Ecal_\Omega^i|$    or $|\Ecal_\Omegabar^i|$.   Let
$N(\Omega)$ be the number  of sites in  the ``bulk'' of $\Omega$, then
$|\Ecal_\Omega^i|$   scales as $\sim   a^{N(\Omega)}$, where  $a$ is a
constant.   Similarly,   $|\Ecal_\Omegabar^i|\sim   a^{N(\Omegabar)}$.
Since we have  $\sim   2^{P(\Omega)}$ possible $\OS_i$'s,  the   total
computation   time adds up to   $\sim  2^{P(\Omega)} ( a^{N(\Omega)} +
a^{N(\Omegabar)} )$.   With the  extension of  the area $\Omega$,  the
number of     possible   $\OS_i$   increases,  but    counting   dimer
configurations  get faster because $\Omegabar$ shrinks.   
Thus this  method is  optimal  for large
areas in medium-size systems  (here up to $N=64$), being complementary
to the Pfaffian  method.  One can reduce the  time further by dividing
$\Omega$    or $\Omegabar$.   Let us   consider   an  annulus like  in
Fig.   \ref{fig:circle_divide_KP} as   $\Omega$,  for  example.  Then
$\Omegabar$  can naturally be divided into  inner ($r<\Rin$) and outer
($r>\Rout$)  parts, denoted by   $\omega$  and $\omega^\prime$.  Since
$\Omega$  has two  disconnected  boundaries,  with $\omega$  and  with
$\omega^\prime$, one can label $\OS (c)$  by two numbers, $i$ and $j$,
corresponding to the  occupations around these boundaries.  The  dimer
configurations  on   $\omega$  and  $\omega^\prime$   can  be  counted
separately for   given $i$ and $j$.  The   eigenvalues to calculate is
therefore   expressed  as  $\lambda_{ij}    =  |\Ecal_\omega^i|  \cdot
|\Ecal_\Omega^{ij}| \cdot |\Ecal_{\omega^\prime}^j|  / |\Ecal|$.   Let
$P$ ($P^\prime$) be the  ``length''  of the boundary  between $\Omega$
and $\omega$ ($\omega^\prime$).  The required computation time becomes
$\sim 2^P\cdot  a^{N(\omega)} +  2^{P+P^\prime}\cdot  a^{N(\Omega)}  +
2^{P^\prime}\cdot a^{N(\omega^\prime)}$. By dividing areas, in general,
one  can reduce the time  of counting configurations  in this way, but
the number of possible  occupations at the  boundaries increases.  One
needs to choose   an efficient division,  depending on  the system and
area sizes.


\begin{thebibliography}{99}

 \bibitem{Srednicki93}
 M. Srednicki, \href{http://link.aps.org/abstract/PRL/v71/e666}{Phys. Rev. Lett. {\bf 71}, 666 (1993)}.
 \bibitem{Vidal03}
 C. Holzhey, F. Larsen, and F. Wilczek, Nucl. Phys. B {\bf 424}, 443 (1994); 
 G. Vidal, J.~I. Latorre, E. Rico and A. Kitaev, \href{http://link.aps.org/abstract/PRL/v90/e227902}{Phys. Rev. Lett. {\bf 90}, 227902 (2003)};
 P. Calabrese and J. Cardy, J. Stat. Mech. (2004) P06002;
 N. Laflorencie, E.~S. Sorensen, M.-S. Chang and I. Affleck, 
 \href{http://link.aps.org/abstract/PRL/v96/e100603}{Phys. Rev. Lett. {\bf 96}, 100603 (2006)}; 
 S. Ryu and T. Takayanagi, \href{http://link.aps.org/abstract/PRL/v96/e181602}{Phys. Rev. Lett. {\bf 96}, 181602 (2006)}.
 \bibitem{Fradkin06}
 E.~Fradkin and  J.~E.~Moore, \href{http://link.aps.org/abstract/PRL/v97/e050404}{Phys. Rev. Lett. {\bf 97}, 050404 (2006)}.
 \bibitem{Wen90}
 X.-G. Wen and Q. Niu, \href{http://link.aps.org/abstract/PRB/v41/e9377}{Phys. Rev. B {\bf 41}, 9377 (1990)}; 
 X.-G. Wen, \href{http://link.aps.org/abstract/PRB/v44/e2664}{Phys. Rev. B, {\bf 44}, 2664 (1991)}.
 \bibitem{Wen04}
 X.-G. Wen, {\it Quantum field theory of many-body systems}, Oxford university press, 2004.
 \bibitem{Ioffe02_PRB}
 L.~B. Ioffe and M.V. Feigel'mann, \href{http://link.aps.org/abstract/PRB/v66/e224503}{Phys. Rev. B {\bf 66}, 224503 (2002)}. 
 \bibitem{Furukawa06}
 S. Furukawa, G. Misguich and M. Oshikawa, \href{http://link.aps.org/abstract/PRL/v96/e047211}{Phys. Rev. Lett. {\bf 96}, 047211 (2006)}; 
 \href{http://www.iop.org/EJ/abstract/0953-8984/19/14/145212}{J. Phys.: Condens. Matter {\bf 19}, 145212 (2007)}.
 \bibitem{Furukawa07}
 S. Furukawa, Ph. D thesis, Tokyo Institute of Technology, 2007.
 \bibitem{Preskill00}
 J. Preskill, J. Mod. Opt. {\bf 47}, 127 (2000).
 \bibitem{Kitaev06}
 A. Kitaev and J. Preskill, \href{http://link.aps.org/abstract/PRL/v96/e110404}{Phys. Rev. Lett. {\bf 96}, 110404 (2006)}.
 \bibitem{Levin06}
 M. Levin and X.-G. Wen, \href{http://link.aps.org/abstract/PRL/v96/e110405}{Phys. Rev. Lett. {\bf 96}, 110405 (2006)}. 
 \bibitem{Moessner01}
 R. Moessner and S. L. Sondhi, \href{http://link.aps.org/abstract/PRL/v86/e1881}{Phys. Rev. Lett. {\bf 86}, 1881 (2001)}.
 \bibitem{Rokhsar88}
 D.~S.~Rokhsar and S.~A.~Kivelson, \href{http://link.aps.org/abstract/PRL/v61/e2376}{Phys. Rev. Lett. {\bf 61}, 2376 (1988)}.
 \bibitem{Kitaev03}
 A. Y. Kitaev, Ann. Phys. (New York) {\bf 303}, 2 (2003).
 \bibitem{Hamma05}
 A. Hamma, R. Ionicioiu and P. Zanardi, Phys. Lett. A {\bf 337}, 22 (2005); 
 \href{http://link.aps.org/abstract/PRA/v71/e022315}{Phys. Rev. A {\bf 71}, 022315 (2005)}.
 \bibitem{Misguich02}
  G.~Misguich, D.~Serban and V.~Pasquier, 
  \href{http://link.aps.org/abstract/PRL/v89/e137202}{Phys. Rev. Lett. {\bf 89}, 137202 (2002)}.
 \bibitem{Haque06}
 M. Haque, O. Zozulya and K. Schoutens, 
  \href{http://link.aps.org/abstract/PRL/v98/e060401}{Phys. Rev. Lett. {\bf 98}, 060401 (2007)}.
 \bibitem{Ioselevich02}
  A. Ioselevich, D.~A.~Ivanov and M.~V.~Feigel'mann, 
  \href{http://link.aps.org/abstract/PRB/v66/e174405}{Phys. Rev. B {\bf 66}, 174405 (2002)}.
 \bibitem{Fendley02}
  P. Fendley, R. Moessner and S. L. Sondhi, 
  \href{http://link.aps.org/abstract/PRB/v66/e214513}{Phys. Rev. B {\bf 66}, 214513 (2002)}.
 \bibitem{Ioffe02_Nature}
 L.~B.~Ioffe, M.~V.~Feigel'man, A.~Ioselevich, D.~Ivanov, M.~Troyer and G.~Blatter, Nature {\bf 415}, 503 (2002).
 \bibitem{Ralko05-06}
 A.~Ralko, M.~Ferrero, F.~Becca, D.~Ivanov and F.~Mila, 
 \href{http://link.aps.org/abstract/PRB/v71/e224109}{Phys. Rev. B {\bf 71}, 224109 (2005)};
 \href{http://link.aps.org/abstract/PRB/v74/e134301}{Phys. Rev. B {\bf 74}, 134301 (2006)}.
 \bibitem{Wegner71}
 F.~J.~Wegner, J. Math. Phys. (N.Y.) {\bf 12}, 2259 (1971).
 \bibitem{Moessner02}
 R. Moessner, S. L. Sondhi and E. Fradkin \href{http://link.aps.org/abstract/PRB/v65/e024504}{Phys. Rev. B {\bf 65}, 024504 (2002)}.
 \bibitem{Nielsen00}
 M. Nielsen and I. L. Chuang, {\it Quantum computation and quantum information}, Cambridge university press, 2000.
 \bibitem{Kasteyleyn61}
  P. W. Kasteyleyn, Physica {\bf 27}, 1209 (1961); J. Math. Phys. {\bf 4}, 287 (1963); 
  see also Refs. \onlinecite{Ioselevich02,Fendley02}.
 \bibitem{Maeda03}
  O. Derzhko and T. Krokhmalskii, Phys. Status Solidi B {\bf 208}, 221 (1998);
  Y. Maeda and M. Oshikawa, \href{http://link.aps.org/abstract/PRB/v67/e224424}{Phys. Rev. B {\bf 67}, 224424 (2003)}.
 
\end{thebibliography}

\end{document}